\documentclass[aps,prd,preprint,showpacs,nofootinbib]{revtex4}
\usepackage{amsmath}
\usepackage{epsfig}
\usepackage{amssymb}

\def\b{\beta}
\def\l{\lambda}
\def\e{\epsilon}
\def\r{\rho}
\def\L{\Lambda}
\def\m{\mu}

\def\G{\Gamma}
\def\f{\phi}
\def\O{\Omega}
\def\a{\alpha}
\def\OP{{\cal O}}

\begin{document}

\title{Evolution of cosmological perturbations in an RG-driven inflationary scenario}

\author{Adriano Contillo}

\affiliation{SISSA, Via Bonomea 265, I-34136 Trieste, Italy\\
INFN, Sezione di Trieste, I-34127 Trieste, Italy}

\begin{abstract}
A gauge-invariant, linear cosmological perturbation theory of an almost homogeneous and
isotropic universe with dynamically evolving Newton constant
G and cosmological constant $\Lambda$ is presented. The
equations governing the evolution of the comoving fractional spatial gradients
of the matter density, G and $\Lambda$ are thus obtained. Explicit solutions are discussed
in cosmologies, featuring an accelerated expansion, where both G and $\Lambda$ vary according to renormalization group equations in the vicinity of an ultraviolet fixed point. Finally, a similar analysis is carried out in the late universe regime described by the part of the renormalization group trajectory close to the gaussian fixed point.
\end{abstract}


\maketitle

\section{Introduction}

The subject of cosmological perturbations continues to attract much attention because
it is an essential step in understanding any theory of cosmological structure formation. The
simple idea that the observed structure in the universe has resulted from the gravitational
amplification of small primordial fluctuations works remarkably well and it must be discussed
in any theory of relativistic cosmology.

In inflationary cosmology the presently observed structures in the universe are generated
by quantum fluctuations during an early de Sitter phase. The subsequent evolution is
classical and depends on the interplay between pressure forces, the rate of growth of the
expansion factor and on the content and the nature of the, yet unknown, dark matter.

Quite recently, an alternative scenario was proposed. 
It is based on the assumption that the gravitational couplings
are subject to renormalization group (RG) flow, and reach a fixed point
when the energy goes to infinity.
Evidence for this has been found first in the Einstein-Hilbert truncation
\cite{reuter1,dou,souma,lauscher,litim,saueressig,eichhorn},
followed by extensions to four-derivative gravity
\cite{codello1,bms,niedermaier}
and to $f(R)$ gravity~\cite{reuter2,cpr,ms,bcp}.
Further extension have been discussed in the conformal truncation,
where one freezes the spin two degrees of freedom
\cite{creh,machado}, 
and in bimetric truncations~\cite{manrique}.

In~\cite{bore1} the RG scenario was applied to a Friedmann-Robertson-Walker cosmological model, assuming that the relevant energy scale upon which the couplings depend is the Hubble parameter $H(t)$. This gave rise to a ``RG improvement'' of Einstein equations that was found to be responsible for a phase of accelerated expansion of the scale factor. RG scenario was thus able to trigger a mechanism of ``inflation'' without the introduction of an \emph{ad hoc} scalar field.

In~\cite{bore2} the evolution of gauge-invariant perturbations, according to the formalism pioneered by Hawking~\cite{ha}, further extended by Ellis and Bruni~\cite{elbr}, Jackson~\cite{ja} and Zimdahl~\cite{zi}, was studied in a homogeneous and isotropic background universe, deep inside the late universe regime (where the authors postulated the existence of an infrared fixed point). It was found that all the relevant perturbations decay with a power law of the cosmic time or stay constant, \emph{i.e.} the IR fixed point cosmology is stable under small deviations from perfect homogeneity and isotropy. The aim of the present paper is to investigate the evolution of cosmological perturbations in the vicinity of the UV fixed point of the RG trajectory, close to the above mentioned accelerated phase of the universe.

It is important to point out the fact that, in~\cite{bore2}, the stress-energy tensor of the perfect fluid filling the universe was assumed to be covariantly conserved, an hypothesis that only works well for low values of the cutoff scale, when the entropy exchange between the matter sector and the coupling constants is suppressed (see~\cite{bore1} for details). Moving backwards along the RG trajectory up to the Planck mass scale, the suppression breaks down and the sudden variation of the cosmological constant $\Lambda$ generates a flow of energy from the geometry into the fluid. As a consequence, the conservation of the stress-energy tensor gets spoiled and one can only count on the more general contracted Bianchi identity of the Einstein tensor.

The paper is structured as follows: in section \ref{background} the main features of the RG improved background cosmology are briefly reported, with a closer look at the very early stages of its evolution. In section \ref{RGperturb} the covariant approach to the evolution of perturbations is described and is applied to the RG-driven inflationary background, getting power law solutions that will be discussed in section \ref{FPsolutions}. Finally, in section \ref{k4regime} the same analysis is applied to the subsequent phase of quasi-classical evolution predicted by the RG improved gravitational theory.

\section{Renormalization Group derived background}\label{background}

The idea of using the RG flow equation in gravity is borrowed from particle physics where the RG improvement is a standard device in order to add, for instance, the dominant quantum corrections to the Born approximation of a scattering cross section. However, instead of improving solutions, in~\cite{bore3,bore4} the more powerful improvement of the basic equations has been discussed. The main results for an homogeneous and isotropic universe will now be briefly reviewed.

It will be assumed that there is a fundamental scale dependence of Einstein-Hilbert coupling constants which is governed by an exact RG equation for a Wilsonian effective action whose precise nature needs not to be specified here. Regardless of the technical details of the chosen renormalization scheme, there is a rather general fact about renormalization group improved actions that is worth to be stressed.
A renormalized action is composed by a sum of operators that depend both on the fields, that we generically call $\f$, and their momenta $p$, coupled to constants depending on a renormalization scale $k$
\begin{equation}
 \G_k=\sum_i g_i(k)\OP_i(\f,p)\;.
\end{equation}
As such scale is typically identified as $k\sim p$, one gets that
\begin{equation}
 \G_k\equiv\sum_i g_i(p)\OP_i(\f,p)=\sum_i\OP'_i(\f,p)
\end{equation}
where the new operators get additional dependences on the momenta, that are rather generic and can even include nonlocality. This means that the renormalization group itself is a kind of shortcut to take into account more terms than the ones actually present in the action. A more exhaustive discussion about this topic can be found in~\cite{rewe1}.

Getting back to the Einstein-Hilbert framework, at a typical length scale $\ell$ or mass scale $k=\ell^{-1}$ the coupling constants assume the values $G(k)$ and $\Lambda(k)$, respectively. In trying to “RG-improve” Einstein’s equation the crucial step is the identification of the scale which is relevant in the situation under consideration. In cosmology, the postulate of homogeneity and isotropy implies that $k$ can depend on the cosmological time only so that the scale dependence is turned into a time dependence:
\begin{equation}\label{gvst}
 G\equiv G(k(t))\;,\;\Lambda\equiv\Lambda(k(t))\;.
\end{equation}

In~\cite{bore1,shaso} this dependence was implemented by setting such energy scale to be proportional to some effective measure of the curvature of the universe, \emph{i.e.} the Hubble rate:
\begin{equation}\label{kvst}
 k(t)=\xi\,H(t)=\xi\,\frac{\dot{a}}{a}
\end{equation}
where $\xi$ is a $O(1)$ parameter that is kept free for the time being. For the sake of completeness, it has to be pointed out that other choices are possible, as for exemple the cutoff $k\sim1/t$ proposed in~\cite{bore3,rewe1,rewe2} or the $k\sim1/a(t)$ in~\cite{flpe,bauer}, while in~\cite{dost,kora,caea} the cutoff dependence was adjusted to fit the energy-momentum conservation of ordinary matter.

Later on, when dealing with perturbations around the Friedmann-Robertson-Walker (FRW) background,
one will have to give up perfect homogeneity and isotropy, so equation (\ref{gvst})
will generalize to
\begin{equation}\label{Gvsx}
 G\equiv G(k(x^\mu))\;,\;\Lambda\equiv\Lambda(k(x^\mu))\;.
\end{equation}
In this context $H$ is not defined anymore, and in section \ref{RGperturb} it will be
identified $k\equiv\xi\,\theta/3$, where $\theta$ is the expansion rate of a congruence of
world lines of the cosmological fluid. In the FRW case $\theta=3H$, so there is consistency with the choice made in (\ref{kvst}). As said before, the cutoff scale is meant here to be identified with some measure of the local curvature, and gets a coordinate dependence only because the curvature happens to be coordinate-dependent. From this point of view, there is no conceptual difference between the choice $k=k(H(t))$ in a FRW universe and the one $k=k(\theta(x^\m))$ in a quasi-FRW one.

For what regards the homogeneous and isotropic background, in the following the main results of~\cite{bore1} are briefly resumed. With $G(k)=g(k)k^{-2}$ and $\Lambda(k)=\lambda(k)k^2$, $g(k)$ and $\lambda(k)$ being the dimensionless couplings, one has
\begin{equation}
 G(t)=\frac{g(\xi\,H(t))}{\xi^2H(t)^2}\;,\;\Lambda(t)=\lambda(\xi\,H(t))\xi^2H(t)^2\;.
\end{equation}
Then, provided $H(t)\neq0$, the cosmological evolution equations for a fluid with
equation of state $p = w\rho$ can be cast in the following form:
\begin{equation}\label{HdotvsH}
 \dot{H}(t)=-\frac{3}{2}(1+w)H(t)^2\left[1-\frac{\xi^2}{3}\lambda(\xi\,H(t))\right]
\end{equation}
\begin{equation}\label{rhovsH}
 \rho(t)=\frac{3\xi^2}{8\pi g(\xi\,H(t))}\left[1-\frac{\xi^2}{3}\lambda(\xi\,H(t))\right]H(t)^4\;.
\end{equation}

In~\cite{bore1} it is stressed that for $k\rightarrow\infty$ the RG flow spirals
onto a Non-Gaussian Fixed Point (NGFP), whose values of the dimensionless coupling
constants are denoted by $g_\ast$ and $\lambda_\ast$. In that paper the spiral is
approximated to a constant-$g$ and $\lambda$ trajectory, so the differential
equation (\ref{HdotvsH}) reads
\begin{equation}\label{FPHdot}
 \dot{H}=-\alpha^{-1}H^2
\end{equation}
with the constant
\begin{equation}
 \alpha\equiv\frac{2}{3(1+w)(1-\lambda_\ast\xi^2/3)}=\frac{2}{3(1+w)(1-\Omega^\ast_{\Lambda})}
\end{equation}
$\Omega^\ast_{\Lambda}$ being the NGFP value of $\Omega_{\Lambda}(\xi
H(t))=\xi^2\lambda(\xi\,H(t))/3$.

Fixing the constant of integration such that this singularity occurs at $t=0$, the
unique solution to (\ref{FPHdot}) reads
\begin{equation}
 H(t)=\frac{\alpha}{t}
\end{equation}
which integrates to $a(t)\propto t^{\alpha}$. Using $\Omega^\ast_{\Lambda}$ as the free
parameter which distinguishes different solutions, the fixed point cosmologies are
characterized by the following power laws:
\begin{equation}\nonumber
 a(t)=a_0t^{\alpha}
\end{equation}
\begin{equation}\nonumber
 \rho(t)=\frac{2\Omega^\ast_{\Lambda}}{9\pi
g_\ast\lambda_\ast(1+w)^4(1-\Omega^\ast_{\Lambda})^3}\;t^{-4}
\end{equation}
\begin{equation}\label{FPeqns}
 G(t)=\frac{3g_\ast\lambda_\ast(1+w)^2(1-\Omega^\ast_{\Lambda})^2}{4 \Omega^\ast_{\Lambda}}\;t^2
\end{equation}
\begin{equation}\nonumber
 \Lambda(t)=\frac{4 \Omega^\ast_{\Lambda}}{3(1+w)^2(1-\Omega^\ast_{\Lambda})^2}\;t^{-2}
\end{equation}
Note that the RG data enter the solution (\ref{FPeqns}) only via the universal
product $g_\ast\lambda_\ast$~\cite{lauscher,saueressig}. By the analysis of the deceleration parameter
\begin{equation}\label{accparam}
 q=\frac{1}{\alpha}-1=\frac{1}{2}[1+3w-3(1+w)\Omega^\ast_{\Lambda}]
\end{equation}
it can be seen that, in a radiation-dominated regime, the universe undergoes an
accelerated expansion if $\Omega^\ast_{\Lambda}>1/2$. In section \ref{FPsolutions} we
compute gauge-invariant perturbations in a NGFP background, which corresponds to the
inflationary phase of the RG improved cosmology.

There is one last thing about the evolution of the background universe that needs to
be defined: after having left the NGFP, the trajectory flows towards the Gaussian Fixed Point
(GFP), passing very close to it. This means that the couplings become small, and then
the flow can be linearized and the $\b$-functions assume the form
\begin{eqnarray}
 \b_g&=&2g\nonumber\\
 \b_\l&=&\frac{g}{2\pi}-2\l
\end{eqnarray}
and the corresponding trajectory can be parametrized as
\begin{eqnarray}\label{ccvsk}
 g(k)&=&4\pi\l_T\left(\frac{k}{k_T}\right)^2=g_T\left(\frac{k}{k_T}\right)^2\nonumber\\
 \l(k)&=&\frac{1}{2}\l_T \left[\left(\frac{k}{k_T}\right)^2+\left(\frac{k_T}{k}\right)^2\right]
\end{eqnarray}
where $\lambda_T\equiv\lambda(k_T)$ and $k_T$ is the value of the cutoff scale for
which
\begin{equation}
 \beta_{\lambda}(k_T)\equiv\left.k\frac{d}{dk}\lambda(k)\right|_{k=k_T}=0
\end{equation}
and that is called \emph{turning point}.

During this epoch, the dimensionful coupling constants behave like
\begin{eqnarray}
 G(k)&=&\bar{G}\nonumber\\
 \L(k)&=&\L_0+\nu\,\bar{G}\,k^4
\end{eqnarray}
where $\nu$ is a constant of order unity whose precise value depends on the renormalization scheme, and at very late times they approach their present observed values $\bar{G}$, $\L_0$. This is the reason why such regime is denominated quasi-classical. These formulas will prove useful in section \ref{k4regime}, when dealing with evolution of perturbations in the late universe.

\section{Gauge-invariant perturbation theory}\label{RGperturb}

In this section a generalization of the formalism of refs.~\cite{ha,elbr,ja,zi} is presented, by allowing $G\equiv G(x^{\mu})$ and
$\Lambda\equiv \Lambda(x^{\mu})$ to be scalar functions on spacetime.

A “fundamental observer” describing the cosmological fluid flow lines has 4-velocity
\begin{equation}
 u^{\mu}=dx^{\mu}/d\tau, \quad u^{\mu}u_{\mu}=-1
\end{equation}
where $\tau$ is the proper time along the flow lines. The projection tensor onto the tangent
3-space orthogonal to $u^{\mu}$ is
\begin{equation}
 h_{\mu\nu}=g_{\mu\nu}+u_{\mu}u_{\nu}
\end{equation}
with $h^{\mu}_{\phantom{\mu}\sigma}h^{\sigma}_{\phantom{\sigma}\nu}=h^{\mu}_{\phantom{\mu}\nu}$ and $h^{\mu}_{\phantom{\mu}\nu}u^{\nu}=0$. The covariant derivative of $u^{\mu}$ is
\begin{equation}\label{udecomp}
 u_{\mu;\nu}=\frac{1}{3}\theta h_{\mu\nu}+\sigma_{\mu\nu}+\omega_{\mu\nu}-\dot{u}_{\mu}u_{\nu}
\end{equation}
where $\sigma_{\mu\nu}=h^{\alpha}_{\phantom{\alpha}\mu}h^{\beta}_{\phantom{\beta}\nu}u_{\left(\alpha;\beta\right)}-\theta h_{\mu\nu}/3$ is the shear tensor, $\omega_{\mu\nu}=h^{\alpha}_{\phantom{\alpha}\mu}h^{\beta}_{\phantom{\beta}\nu}u_{\left[\alpha;\beta\right]}$ is the vorticity tensor, $\theta=u^{\mu}_{\phantom{\mu};\mu}$ is the expansion scalar and $\dot{u}^{\mu}=u^{\mu}_{\phantom{\mu};\nu}u^{\nu}$ is the acceleration four-vector (square and round brackets denote anti-symmetrization and symmetrization, respectively). The Riemann tensor is defined by $u_{\mu;\rho\sigma}-u_{\mu;\sigma\rho}=R^{\phantom{\rho\sigma\mu}\lambda}_{\rho\sigma\mu}u_{\lambda}$ and the Ricci tensor $R_{\mu\nu}=R^{\tau}_{\phantom{\tau}\mu\tau\nu}$. With these conventions, Einstein’ s equations are
\begin{equation}
 G_{\mu\nu}=R_{\mu\nu}-\frac{1}{2}Rg_{\mu\nu}=8\pi GT_{\mu\nu}-\Lambda g_{\mu\nu}
\end{equation}
where $\Lambda=\Lambda(x^{\mu})$ is the coordinate-dependent cosmological constant and $G=G(x^{\mu})$ the
coordinate-dependent Newton constant. As already said, the conservation constraint for energy-momentum tensor $T^{\mu\nu}_{\phantom{\mu\nu};\nu}=0$ is going to be dropped in favour of the less restrictive hypothesis
\begin{equation}\label{bianchi}
 G^{\mu\nu}_{\phantom{\mu\nu};\nu}=\left( 8\pi GT^{\mu\nu}-\Lambda g^{\mu\nu} \right)_{;\nu}=0\;.
\end{equation}

A perfect fluid stress-energy tensor has the form $T^{\mu\nu}=\rho u^{\mu}u^{\nu}+ph^{\mu\nu}$, so eqn. (\ref{bianchi}) leads to a modified energy conservation
\begin{equation}\label{conservation}
\dot{\rho}+ \theta (\rho +p)+\frac{\dot{G}}{G} \rho +\frac{\dot{\Lambda}}{8 \pi G}=0
\end{equation}
and a modified equation of motion
\begin{equation}\label{motion}
\dot{u}^{\mu}+\frac{h^{\mu\nu}}{\rho +p}\left(p_{;\nu}+\frac{G_{;\nu}}{G} p-\frac{\Lambda_{;\nu}}{8\pi G}\right)=0
\end{equation}
by projecting along $u^{\mu}$ and onto the hyperplane orthogonal to $u^{\mu}$.

By a redefinition of the variables it is possible to rearrange the previous equations in a shape that is formally identical to the usual stress-energy conservation: defining the quantities
\begin{equation}
{\cal R} \equiv G \rho +\frac{\Lambda}{8 \pi}\; , \quad {\cal P} \equiv G p -\frac{\Lambda}{8 \pi}
\end{equation}
equations (\ref{conservation}) and (\ref{motion}) get recast in
\begin{equation}
\dot{{\cal R}}+ \theta ({\cal R} +{\cal P})=0
\end{equation}
and
\begin{equation}
\dot{u^{\mu}}+\frac{h^{\mu\nu}{\cal P}_{;\nu}}{{\cal R}+{\cal P}}=0
\end{equation}
with the new variables playing the role of some effective $\rho$ and $p$.

Combining the Einstein equations and the (\ref{udecomp}) one gets the well known Raychaudhuri equation, that in the language of $\cal R$ and $\cal P$ is written as
\begin{equation}\label{ray}
 \dot{\theta}+\frac{1}{3}\theta^2+2(\sigma^2-\omega^2)-\dot{u}^{\mu}_{\phantom{\mu};\mu}+4\pi({\cal R}+3{\cal P})=0
\end{equation}
where $2\sigma^2\equiv \sigma_{\mu\nu}\sigma^{\mu\nu}$ and $2\omega^2\equiv \omega_{\mu\nu}\omega^{\mu\nu}$. The term $\dot{u}^{\mu}_{\phantom{\mu};\mu}$ can be rewritten as~\cite{zi}
\begin{equation}
 \dot{u}^{\mu}_{\phantom{\mu};\mu}=-h^{\lambda\rho}\left(\frac{h^{\nu}_{\phantom{\nu}\lambda}{\cal P}_{;\nu}}{{\cal R}+{\cal P}}\right)+h^{\lambda\rho}\frac{{\cal P}_{;\lambda}}{{\cal R}+{\cal P}}\,\frac{{\cal P}_{;\rho}}{{\cal R}+{\cal P}}\;,
\end{equation}
the scalar $\cal K$ is defined as
\begin{equation}
 {\cal K}\equiv 2\sigma^2-\frac{2}{3}\theta^2+16\pi{\cal R}
\end{equation}
and it is possible to show that for zero vorticity ($\omega_{\mu\nu}=0$) it coincides with the Ricci scalar $^{(3)}\!R$ of the 3-dimensional hyperplane everywhere orthogonal to $u^{\mu}$. An auxiliary length scale $S(t)$
is introduced as the solution of the equation
\begin{equation}
 \frac{\dot{S}}{S}=\frac{\theta}{3}\;.
\end{equation}

Suitable quantities useful to characterize the spatial inhomogeneities of density, pressure
and expansion should be, respectively,
\begin{equation}\label{defD}
 D_{\mu}\equiv\frac{Sh^{\nu}_{\phantom{\nu}\mu}\,\rho_{;\nu}}{\rho+p}\;,\quad P_{\mu}\equiv\frac{Sh^{\nu}_{\phantom{\nu}\mu}\,p_{;\nu}}{\rho+p}\;,\quad t_{\mu}\equiv Sh^{\nu}_{\phantom{\nu}\mu}\theta_{;\nu}
\end{equation}
and to characterize spatial gradients of $G$ and $\Lambda$ can be introduced the dimensionless quantities
\begin{equation}\label{defGamma} \Gamma_{\mu}\equiv\frac{Sh^{\nu}_{\phantom{\nu}\mu}G_{;\nu}}{G}\;,\quad\Delta_{\mu}\equiv\frac{Sh^{\nu}_{\phantom{\nu}\mu}\Lambda_{;\nu}}{\Lambda}
\end{equation}
as it is already done in~\cite{bore2}. However, as in the present case we are working with some adjusted density and pressure, the most useful quantities turn out to be
\begin{equation}\label{defcalD}
 {\cal D}_{\mu}\equiv\frac{Sh^{\nu}_{\phantom{\nu}\mu}{\cal R}_{;\nu}}{{\cal R}+{\cal P}}\;,\quad\varPi_{\mu}\equiv\frac{Sh^{\nu}_{\phantom{\nu}\mu}{\cal P}_{;\nu}}{{\cal R}+{\cal P}}\;.
\end{equation}

One can now write, using (\ref{udecomp}), (\ref{conservation}) and (\ref{motion}),
\begin{equation}\nonumber
 h^{\nu}_{\phantom{\nu}\mu}(Sh^{\lambda}_{\phantom{\lambda}\nu}{\cal R}_{;\lambda})\dot{\,}=\frac{S}{3}\theta h^{\nu}_{\phantom{\nu}\mu}{\cal R}_{;\nu}+Sh^{\nu}_{\phantom{\nu}\mu}{\cal P}_{;\nu}+Sh^{\nu}_{\phantom{\nu}\mu}({\cal R}_{;\nu})\dot{\,}=
\end{equation}
\begin{equation}\label{calDdot}
 =-S\theta h^{\nu}_{\phantom{\nu}\mu}{\cal R}_{;\nu}-Sh^{\nu}_{\phantom{\nu}\mu}\theta_{;\nu}({\cal R}+{\cal P})-Sh^{\nu}_{\phantom{\nu}\lambda}{\cal R}_{;\nu}(\sigma^{\lambda}_{\phantom{\lambda}\mu}+\omega^{\lambda}_{\phantom{\lambda}\mu})
\end{equation}
where in the second line the fact was used that $({\cal R}_{;\mu})\dot{\,}=\dot{\cal R}_{;\mu}-u^{\nu}_{\phantom{\mu};\mu}{\cal R}_{;\nu}$.

The expression (\ref{calDdot}) can be rewritten as a differential equation involving the adjusted gradients (\ref{defcalD})
\begin{equation}\label{Ddot}
h^{\nu}_{\phantom{\nu}\mu}\dot{{\cal D}}_{\nu}+\frac{\dot{{\cal P}}}{{\cal R}+{\cal P}}{\cal D}_{\mu}+ (\omega^{\nu}_{\phantom{\nu}\mu}+\sigma^{\nu}_{\phantom{\nu}\mu}){\cal D}_{\nu}+t_{\mu}=0
\end{equation}
and following a similar procedure one can also obtain an equation for $t_{\mu}$
\begin{equation}\nonumber
 h^{\nu}_{\phantom{\nu}\mu}\dot{t}_{\nu}+\dot{\theta}\varPi_{\mu}+\frac{2}{3}\theta t_{\mu}+(\omega^{\nu}_{\phantom{\nu}\mu}+\sigma^{\nu}_{\phantom{\nu}\mu})t_{\nu}+2Sh^{\nu}_{\phantom{\nu}\mu}(\sigma^2-\omega^2)_{;\mu}+
\end{equation}
\begin{equation}\label{tdot}
-Sh^{\lambda}_{\phantom{\lambda}\mu}(\dot{u}^{\nu}_{\phantom{\nu};\nu})_{;\lambda}+4\pi({\cal R}+{\cal P})({\cal D}_{\mu}+3\varPi_{\mu})=0
\end{equation}
where Raychaudhuri equation (\ref{ray}) was used to extract the final expression.

Notice that both (\ref{Ddot}) and (\ref{tdot}) represent the most general extension of the usual gauge-invariant perturbation theory equations in the case of non-constant $G$ and $\Lambda$, in which no assumptions are made on their functional form.

From now on we will focus on a homogeneous and isotropic background universe, for which $\sigma_{\mu\nu}=\omega_{\mu\nu}=\dot{u}_{\mu}=0$, $S=a$, $^{(3)}\!R =6K/a^2$ and $\theta=3H$. In such universe the relevant equations for the background evolution read
\begin{equation}
 {\cal K}=2(-\frac{1}{3}\theta^2+8\pi{\cal R})=^{(3)}\!\!\!R
\end{equation}
\begin{equation}
 \dot{\theta}+12\pi({\cal R}+{\cal P})=^{(3)}\!\!\!R/2\;.
\end{equation}

We now consider small perturbations of the motion of the fluid and consequently, up to first order in the inhomogeneities, the factors in the equations multiplying the quantities ${\cal D}_{\mu}$, $\varPi_{\mu}$ and $t_{\mu}$ refer to the background. Under these hypothesis (\ref{Ddot}) and (\ref{tdot}) get the following form:
\begin{equation}\label{DdotFRW}
 h^{\nu}_{\phantom{\nu}\mu}\dot{{\cal D}}_{\nu}+\frac{\dot{{\cal P}}}{{\cal R}+{\cal P}}{\cal D}_{\mu}+t_{\mu}=0
\end{equation}
\begin{equation}\label{tdotFRW}
 h^{\nu}_{\phantom{\nu}\mu}\dot{t}_{\nu}+\frac{3K}{a^2}\varPi_{\mu}+2Ht_{\mu}+4\pi({\cal R}+{\cal P}){\cal D}_{\mu}+\frac{\nabla^2}{a^2}\varPi_{\mu}=0
\end{equation}
where the fact that~\cite{zi}, up to linear order, $Sh^{\lambda}_{\phantom{\lambda}\mu}(\dot{u}^{\nu}_{\phantom{\nu};\nu})_{;\lambda}\simeq-\nabla^2\varPi_{\mu}/a^2$  was used ($\nabla^2$ is the laplacian on the 3-hypersurface).

For the sake of semplicity, from now on we will focus on a spatially flat FRW, setting $K=0$. One also would like to assign to the perturbed quantities an equation of state $p(\rho)=w\rho$ but, while this translates directly in a relation $P_{\mu}=wD_{\mu}$, it doesn't happen the same for ${\cal D}_{\mu}$ and $\varPi_{\mu}$. A possible solution is to rewrite the latter quantities in terms of the formers
\begin{equation}\label{calDvsD}
 {\cal D}_{\mu}=D_{\mu}+\Gamma_{\mu}\frac{\rho}{\rho+p}+\frac{\Lambda}{8\pi G}\frac{\Delta_{\mu}}{\rho+p}
\end{equation}
\begin{equation}\label{PivsP}
 \varPi_{\mu}=P_{\mu}+\Gamma_{\mu}\frac{\rho}{\rho+p}-\frac{\Lambda}{8\pi G}\frac{\Delta_{\mu}}{\rho+p}
\end{equation}
and then apply the usual equation of state.

There is one last missing ingredient to the recipe: in order to end up with a consistent set of equations for spatial perturbations, it is necessary to eliminate the variables $\Gamma_{\mu}$ and $\Delta_{\mu}$ in favour of the remaining ones. This is achieved by making explicit use of the cutoff identification. It is assumed that the cutoff $k$ appearing in equation (\ref{Gvsx}) is $k=\xi\,\theta/3$ (which reduces to $k=\xi\,H$ in the FRW background). Then, using equations (\ref{defD}) and (\ref{defGamma}), as long as the trajectory keeps close enough to the NGFP, where $g\simeq g_\ast$ and $\lambda\simeq\lambda_\ast$, one can write
\begin{equation}
 \Gamma_{\mu}=\frac{Sh^{\nu}_{\phantom{\nu}\mu}G_{;\nu}}{G}=\frac{Sh^{\nu}_{\phantom{\nu}\mu}((\theta/3)^{-2}g)_{;\nu}}{(\theta/3)^{-2}g}\simeq-\frac{2}{\theta}Sh^{\nu}_{\phantom{\nu}\mu}\theta_{;\nu}=-\frac{2}{3H}t_{\mu}
\end{equation}
\begin{equation}
 \Delta_{\mu}=\frac{Sh^{\nu}_{\phantom{\nu}\mu}\Lambda_{;\nu}}{\Lambda}=\frac{Sh^{\nu}_{\phantom{\nu}\mu}((\theta/3)^2\lambda)_{;\nu}}{(\theta/3)^2\lambda}\simeq\frac{2}{\theta}Sh^{\nu}_{\phantom{\nu}\mu}\theta_{;\nu}=\frac{2}{3H}t_{\mu}
\end{equation}
where in the last step one could get back to $H$, at least up to first order, because it is multiplied to the perturbed quantity $t_{\mu}$. To better understand the meaning of what just done it is worth stressing the fact that, by including $\Gamma_{\mu}$ and $\Delta_{\mu}$ inside the equations, one was treating spatial fluctuations of the coupling constants as some sort of matter fields, whose dynamics were governed by the Einstein equations. But the dynamics of $G$ and $\Lambda$ are described by the RG equations and, by means of the cutoff identification, depend on the expansion rate $\theta$. This means the spatial gradients of the couplings turn out to the proportional to the gradient $t_{\mu}$ of the expansion rate.

It is now time to plug all the ingredients inside the system, ending up with the two differential equations
\begin{eqnarray}\nonumber
 h^{\nu}_{\phantom{\nu}\mu}\dot{D}_{\nu}+\frac{\Lambda-8\pi G\rho}{12\pi G(1+w)\rho H}\,h^{\nu}_{\phantom{\nu}\mu}\dot{t}_{\nu}+\frac{8\pi w(\dot{G}\rho+G\dot{\rho})-\dot{\Lambda}}{8\pi G(1+w)\rho}\,D_{\mu}+\\\nonumber
 -\frac{1}{6(4\pi G(1+w)\rho H)^2}\left[H\Lambda(8\pi\rho\dot{G}+\dot{\Lambda})+\right.\\\label{1stord1}
 +32\pi^2G^2\rho\left((1+w)\rho(3(1+w)H^2+2\dot{H})-2wH\dot{\rho}\right)+\\\nonumber
 \left.-8\pi G\left((1+w)\Lambda\rho\dot{H}+H(8\pi w\rho^2\dot{G}-(2+w)\rho\dot{\Lambda}+\Lambda\dot{\rho})\right)\right]\,t_{\mu}=0
\end{eqnarray}
\begin{eqnarray}\nonumber
 h^{\nu}_{\phantom{\nu}\mu}\dot{t}_{\nu}+\left(4\pi G(1+w)\rho+w\frac{\nabla^2}{a^2}\right)D_{\mu}+\frac{1}{12\pi G(1+w)\rho H}\Big[8\pi G\rho\\\label{1stord2}
 \left(3(1+w)H^2-4\pi G(1+w)\rho-\frac{\nabla^2}{a^2}\right)+\Lambda\left(4\pi G(1+w)\rho-\frac{\nabla^2}{a^2}\right)\Big]\,t_{\mu}=0
\end{eqnarray}
and, having noticed that $h^{\mu}_{\phantom{\mu}\lambda}\left(h^{\nu}_{\phantom{\nu}\mu}D_{\nu}\right)\dot{}=h^{\nu}_{\phantom{\nu}\lambda}\ddot{D}_{\nu}-\dot{u}_{\lambda}\dot{u}^{\nu}D_{\nu}=h^{\nu}_{\phantom{\nu}\lambda}\ddot{D}_{\nu}$, one can combine them to form a single, second order differential equation in $D_{\mu}$ that for the time being will be schematically indicated as
\begin{equation}\label{2ndord}
 {\cal A}(t)\,h^{\nu}_{\phantom{\nu}\mu}\ddot{D}_{\nu}+{\cal B}(t)\,h^{\nu}_{\phantom{\nu}\mu}\dot{D}_{\nu}+{\cal C}(t)\,D_{\mu}=0\;.
\end{equation}

\section{Fixed Point evolution of perturbations}\label{FPsolutions}

In the following the general formalism developed above is applied to the NGFP background cosmology described in section \ref{background}. First one needs to expand the perturbed quantities in spherical harmonics, that is
\begin{equation}
D^{\mu}=\phi^n(t)\Psi^{\mu}_n
\end{equation}
where $\Psi^{\mu}_n$ is the eigenfunction of the spatial Laplacian operator $\nabla^2$ with eigenvalues $-\epsilon^2_n$. For every $n$ then equation (\ref{2ndord}) becomes
\begin{equation}\label{2ndordsh}
 {\cal A}(t)\,\ddot{\phi}^n(t)+{\cal B}(t)\,\dot{\phi}^n(t)+{\cal C}(t)\,\phi^n(t)=0\
\end{equation}
where, imposing the background (\ref{FPeqns})
\begin{eqnarray}
 {\cal A}(t)\!&=&\!-\frac{3(1+w)}{6w(1-\O_\L^\ast)+2(1+\O_\L^\ast)-3\epsilon_n^2t^2(1+w)(2\O_\L^\ast-1)}\\
 {\cal B}(t)\!&=&\!\frac{t^{-1}}{(\O_\L^\ast-1)\left(6w(\O_\L^\ast-1)-2(\O_\L^\ast+1)+3\epsilon_n^2t^2(1+w)(2\O_\L^\ast-1)\right)^2}\nonumber\\
 &&\times\left[4(1-3w(-1+\O_\L^\ast)+\O_\L^\ast)(2+3w(\O_\L^\ast-1)+3\O_\L^\ast)+\right.\nonumber\\
 &&+\left.9\epsilon_n^4t^4(1+w)^2(\O_\L^\ast-1)(2\O_\L^\ast+1)(1+w(2\O_\L^\ast-1))+\right.\nonumber\\
 &&+\left.6\epsilon_n^2t^2(1+w)((8-13\O_\L^\ast)\O_\L^\ast+3w^2(\O_\L^\ast-1)^2(2\O_\L^\ast-1)+\right.\nonumber\\
 &&\left.-2w(\O_\L^\ast-1)(-2+\O_\L^\ast(5+\O_\L^\ast)))\right]\nonumber\\
 {\cal C}(t)\!&=&\!\frac{t^{-2}}{(1+w)(\O_\L^\ast-1)^2(6w(\O_\L^\ast-1)-2(\O_\L^\ast+1)+3\epsilon_n^2t^2(1+w)(2\O_\L^\ast-1))^2}\nonumber\\
 &&\times\left[4(-1+3w(\O_\L^\ast-1)-\O_\L^\ast)(1+w(\O_\L^\ast-1)+\O_\L^\ast)(-1+\right.\nonumber\\
 &&\left.+3w(\O_\L^\ast-1)+3\O_\L^\ast)-9\epsilon_n^4t^4(1+w)^2(\O_\L^\ast-1)(2\O_\L^\ast-1)(2\O_\L^\ast+\right.\nonumber\\
 &&\left.+w(-1+w(\O_\L^\ast-1)(4\O_\L^\ast-3)+\O_\L^\ast(4\O_\L^\ast-1)))+\right.\nonumber\\
 &&\left.+6\epsilon_n^2t^2(1+w)(3w^3(\O_\L^\ast-1)^3+w^2(\O_\L^\ast-1
)^2(\O_\L^\ast-1)+\right.\nonumber\\
 &&\left.+(\O_\L^\ast+1)(3\O_\L^\ast-1)(4\O_\L^\ast-3)+w(1+\O_\L^\ast(9+\O_\L^\ast(27\O_\L^\ast-37))))\right]\;.\nonumber
\end{eqnarray}

The general solution of (\ref{2ndordsh}) can only be obtained by numerical analysis. Nevertheless, it admits an analytical solution in the form of a power-law $\phi(t)=\Phi\,t^p$ in the long wavelength limit $\epsilon_n\rightarrow0$. By direct substitution, one can find the two solutions
\begin{equation}\label{exponents}
p_-=1-\frac{2}{(1+w)(1-\Omega^\ast_{\Lambda})}\;,\quad p_+=2-\frac{4}{3(1+w)(1-\Omega^\ast_{\Lambda})}
\end{equation}
that do not depend on the values of $\lambda_\ast$ and $g_\ast$, but only on the physical quantity $\Omega^\ast_{\Lambda}$, as it was expected.

While the value of the mode $p_-$ is always negative (remember that in a spatially flat Friedmann-Robertson-Walker $0\leq\Omega^\ast_{\Lambda}\leq1$), $p_+$ can be either a growing or a decreasing mode. In particular, it changes sign from positive to negative for
\begin{equation}
 \Omega^\ast_{\Lambda}=1-\frac{2}{3(1+w)}
\end{equation}
that becomes $\Omega^\ast_{\Lambda}=1/3$ in a matter-dominated regime and $\Omega^\ast_{\Lambda}=1/2$ in radiation-dominated one. This happens to be the same range for which the deceleration parameter $q$ is negative (see section \ref{background}). This means that in the present framework an accelerated expansion of the background, with $q<0$ and thereby $\alpha>1$, corresponds to a decreasing of the perturbations ($p_\pm<0$), as expected in a standard inflationary scenario.

The argument just described is a surprising similarity between the outcomes of the RG improved cosmology and those of the well-known inflationary theories based on the existence of an \emph{inflaton} field~\cite{guth,linde,lyri}, characterized by a non-vanishing Vacuum Expectation Value that plays the role of a vacuum energy. Adding this up to what already presented in~\cite{bore1}, one can conclude that up to now RG improved cosmology is a self-consistent alternative to the usual inflationary paradigm.

\section{Quasi-classical evolution of perturbations}\label{k4regime}

It is now worth checking if the RG improved cosmological perturbations show a late time behaviour that is in agreement with the usual $\L$CDM scenario. In order to achieve this, one has to apply the evolution equation (\ref{2ndord}) to the background described at the end of section \ref{background}. The starting point is the trajectory descibed in equation (\ref{ccvsk}), that can be put in the closed form
\begin{equation}
 \l(g)=\frac{1}{8\pi}\left(g+\frac{g_T^2}{g}\right)
\end{equation}
which can be inverted to give the relation
\begin{equation}\label{gvsl}
 g(\l)=4\pi\left(\l+\e\sqrt{\l^2-\l_T^2}\right)
\end{equation}
where $\e=\pm1$ in the branch of the trajectory where $k\gtrless k_T$.

For what regards the Hubble rate, its evolution equation with respect to $\l$ is
\begin{equation}\label{Hprime}
 \frac{d}{d\l}H(\l)=\frac{H(\l)}{\b_\l}
\end{equation}
and for the $\b$-function one can write
\begin{equation}
 \b_\l(\l)=\frac{g(\l)}{2\pi}-2\l=2\e\sqrt{\l^2-\l_T^2}
\end{equation}
so that (\ref{Hprime}) admits the solution
\begin{equation}\label{Hvsl}
 H(\l)=H_T\left(\frac{\l+\sqrt{\l^2-\l_T^2}}{\l_T}\right)^{1/2\e}
\end{equation}
where $H_T\equiv H(\l_T)$.

The last quantity that needs to be computed is the matter energy density
\begin{equation}
 \r=\frac{3}{8\pi G}\left(H^2-\L/3\right)
\end{equation}
that using (\ref{gvsl}) and (\ref{Hvsl}) becomes
\begin{equation}
 \r(\l)=\frac{3H_T^4}{32\pi^2}\,\xi^2\left(1-\frac{\xi^2\l}{3}\right)\frac{\left(\frac{\l+\sqrt{\l^2-\l_T^2}}{\l_T}\right)^{2/\e}}{\l+\e\sqrt{\l^2-\l_T^2}}\;.
\end{equation}

One can simplify the expressions above obtained using the variable $z=\l/\l_T$, getting for the dimensionless Newton constant
\begin{equation}
 g(z)=4\pi\l_T\left(z+\e\sqrt{z^2-1}\right)
\end{equation}
for the Hubble rate
\begin{equation}
 H(z)=H_T\left(z+\sqrt{z^2-1}\right)^{1/2\e}
\end{equation}
and for the matter energy density
\begin{equation}
 \r(z)=\frac{3H_T}{32\pi^2\l_T}\,\xi^2\left(1-\frac{\xi^2\l_T}{3}z\right)\frac{\left(z+\sqrt{z^2-1}\right)^{2/\e}}{z+\e\sqrt{z^2-1}}\;.
\end{equation}

Plugging the above results into equation (\ref{2ndord}) one ends up with the following second order differential equation for the gradient $D_\m$
\begin{equation}\label{Deqn}
 A\,\ddot{D}_\m+B\,\dot{D}_\m+C\,D_\m=0
\end{equation}
where $A$, $B$ and $C$ are $z$-dependent coefficients. In order to solve the (\ref{Deqn}) in terms of the new variable $z$, one must keep in mind that, given a function $f$
\begin{eqnarray}\label{chainrule}
 \frac{df}{dt}&=&\frac{df}{dz}\,\frac{dz}{d\l}\,\frac{d\l}{dk}\,\frac{dk}{dt}=\frac{df}{dz}\,\frac{1}{\l_T}\b_\l\frac{1}{k}\,\frac{dk}{dt}=\frac{df}{dz}\,\frac{1}{\l_T}\,\b_\l\frac{1}{H}\,\frac{dH}{dt}=\nonumber\\
 &=&\frac{df}{dz}\,\frac{1}{\l_T}\left(\frac{g}{2\pi}-2\l\right)\left(-\frac{3}{2}(1+w)H\left(1-\frac{\xi^2\l}{3}\right)\right)=\\
&=&\frac{3}{2}(1+w)\left(1-\frac{\xi^2\l_T}{3}z\right)\left(2z-\frac{g(z)}{2\pi\l_T}\right)H(z)\frac{df}{dz}\nonumber
\end{eqnarray}
so that, being $\f$ the harmonic expansion of $D_\m=\f^n \Psi_\m^n$, equation (\ref{Deqn}) becomes
\begin{equation}\label{2ordphi}
 \tilde{C}_2(z)\,\f''(z)+\tilde{C}_1(z)\,f'(z)+\tilde{C}_0(z)\,\f(z)=0
\end{equation}
where the tilde was used to underline the fact that, due to (\ref{chainrule}), the coefficients have changed.

Equation (\ref{2ordphi}) admits for $z\gg1$ an analytical solution in the form of a power law. In this limit, the coefficients assume the form
\begin{eqnarray}
 &&\tilde{C}_2(z)\simeq\frac{9(1+w)^3\O_T}{1+4\e+3w(1+2\e)}\,z\nonumber\\
 &&\tilde{C}_1(z)\simeq\frac{9(1+w)^3\O_T\e(3+4\e)}{(1+\e)(1+4\e+3w(1+2\e))}\,z^2\\
 &&\tilde{C}_0(z)\simeq\frac{18(1+w)^3\O_T\e^2}{(1+\e)(1+4\e+3w(1+2\e))}\,z^3
\end{eqnarray}
where $\O_T\equiv\xi^2\l_T/3$. Writing $\f=Az^\a$, one finds for the exponent the ($w$-independent) values
\begin{equation}
 \a=-\frac{1}{2\e}\qquad\textrm{and}\qquad\a=-\frac{1+\e}{\e}
\end{equation}
that for the positive branch (i.e. $k>k_T$) means $\a_-=-2$ and $\a_+=-1/2$, while for the negative one means $\a_-=0$ and $\a_+=1/2$. For what regards the positive branch, one has to make clear that the dimensionless cosmological constant is decreasing with respect to the cosmological time, so the perturbations increase. The negative branch results are even more interesting, because they describe the evolution of the perturbations during the quasi-classical regime of the RG improved cosmology. In the limit for $k\ll k_T$ the cosmological constant can be neglected and the background can be approximated to a standard Einstein-Hilbert (EH) cosmology, so the evolution of its scale factor can be described to a good accuracy as
\begin{equation}
 a(t)=a_0 t^{\frac{2}{3(1+w)}}
\end{equation}
so that inverting the relation gives
\begin{equation}
 t(a)=\left(\frac{a}{a_0}\right)^{\frac{3(1+w)}{2}}
\end{equation}
and beacuse
\begin{equation}
 H(t)=\frac{2}{3(1+w)}\,t^{-1}
\end{equation}
one ends up with the result
\begin{equation}
 H(a)=\frac{2}{3(1+w)}\left(\frac{a}{a_0}\right)^{-\frac{3(1+w)}{2}}
\end{equation}
but from (\ref{ccvsk}) it is known that for $k\ll k_T$
\begin{equation}
 z=\frac{\l}{\l_T}\simeq\frac{1}{2}\frac{k_T^2}{k^2}=\frac{1}{2}\frac{H_T^2}{H^2}
\end{equation}
so that it can be written
\begin{equation}
 z(a)=\frac{9(1+w)^2}{8}H_T^2\left(\frac{a}{a_0}\right)^{3(1+w)}
\end{equation}
and then the growing mode of the density perturbation goes like
\begin{equation}
 \f_\text{grow}(a)=\f_0\left(\frac{a}{a_0}\right)^{3(1+w)/2}\;.
\end{equation}

For a radiation-dominated universe one has that the growing mode $\f_\text{grow}$ goes like $a^2$, while in a matter-dominated one it goes like $a^{3/2}$. This is only partially in agreement with the standard theory proposed in~\cite{bard}: from there the evolutions were found to be $\propto\!\!a^2$ and $\propto\!\!a$ respectively, so that only the radiation-dominated regime coincides with the present analysis. Nonetheless, one has to keep in mind that the present realization of the renormalization group is based on an averaging procedure over momentum scales greater than $H$, so that it only makes sense to give a position dependence to the coupling constants when they are evaluated over distances $\ell\gtrsim H^{-1}$. All the modes lying inside the curvature radius should therefore be evaluated at the averaged cutoff scale $H$, so that the spatial gradients $\Gamma_\mu$ and $\Delta_\mu$ defined in (\ref{defGamma}) should automatically vanish. This is true in particular for the very late stages of the universe, when the modes of the perturbations that lie inside the observable window reenter the horizon. Such statement is consistent with the long wavelenght approximation used throughout the paper, as for spatial momenta greater than the Hubble rate the formalism should break down and a switch to the standard evolution should be mandatory.

\section{Conclusions}

The analysis performed, by means of the gauge-invariant cosmological perturbation theory, provided a description of the evolution of the energy density fluctuations around a background where the coupling constants vary according to renormalization group equations.

Analytical solutions were found only for perturbations with very long wavelength compared to the typical curvature radius of the backgrounf universe. An intriguing issue would be to study the wavelength dependence, in order to construct a power spectrum of the primordial cosmological perturbations. This will be the aim of a future communication.

A special thank to Roberto Percacci, Carlo Baccigalupi and Alfio Bonanno for useful suggestions and comments.

\end{document}